\documentclass[twocolumn,english,aps,pra,showpacs,superscriptaddress]{revtex4}
\usepackage{pslatex}
\usepackage[T1]{fontenc}
\usepackage[latin1]{inputenc}
\usepackage{amssymb}

\makeatletter

\providecommand{\LyX}{L\kern-.1667em\lower.25em\hbox{Y}\kern-.125emX\@}

\usepackage{babel}
\usepackage{graphicx}
\makeatother
\input{epsf}
\begin{document}

\title{Coupled-channel pseudo-potential description
of the Feshbach resonance in two dimensions}

\author{K. Kanjilal and D. Blume}
\affiliation{Department of Physics and Astronomy, Washington State
University, Pullman, WA 99164-2814, USA}

\begin{abstract}
We derive pseudo-potentials that describe 
the scattering between two particles in two spatial dimensions for any
partial wave $m$, whose scattering strength is parameterized in
terms of the phase shift $\delta_{m}$. 
Using our $m=0$ pseudo-potential, we develop
a coupled channel model with 2D zero-range interactions,
which describes the two-body physics across a Feshbach resonance.
Our model predicts the scattering length, the 
binding energy and the ``closed channel molecular fraction''
of two particles; these
observables
can be measured in experiments
on ultracold quasi-2D atomic Bose and Fermi gases
with present-day technology.
\end{abstract}


\maketitle

The success in creating Bose-Einstein condensates 
and Fermi degenerate gases has resulted in a renewed interest 
in the scattering between two atoms.
At very low temperatures the de Broglie wavelengths of the atoms are large
compared to the typical Van der Waals length of atom-atom potentials.
As a result, atoms at ultracold temperatures do not probe the 
detailed structure of the interaction potential. Consequently, the
shape-dependent potential can be replaced
by an appropriate zero-range (ZR) 
pseudo-potential, which is characterized by a 
{\em{single parameter}}, the generalized scattering length.
For example, $s$-wave scattering in three
spatial dimensions
can in many cases be described accurately through Fermi-Huang's 
pseudo-potential~\cite{huan57}.
Generalizations of Fermi-Huang's pseudo-potential, applicable
to higher partial wave scattering in 3D, have also been
considered~\cite{huan57,kanj04,stoc05}.
The foremost advantage of using 
ZR pseudo-potentials is that a number of two-body 
and some many-body problems
become analytically solvable, thus 
highlighting the physical meaning of
a few key parameters.

An exciting development in the area of ultracold atom physics
is the creation of low-dimensional quantum gases.
The trapping of bosons in quasi-1D geometries, e.g., has
allowed the fermionization of 1D Bose gases to be verified 
experimentally~\cite{bloch04} and the
binding energy of 1D molecules 
to be extracted from experiment~\cite{mori05}.
This paper considers quasi-2D systems, for which the motion
in the tight confining direction, the $z$-direction, is frozen 
out~\cite{petr00,grimm}.
The equation of state then depends on the 
(generalized)
2D scattering length.
A thorough understanding of the two-particle physics  
in 2D for any partial wave would aid studies of
the 2D many-body problem. 
Toward this end, we derive 2D ZR pseudo-potentials for any partial
wave $m$ and develop a coupled channel model applicable
to two particles with $m=0$ 
across a Feshbach resonance.

Experiments on ultracold gases routinely utilize 
Feshbach resonances, which allow the
interaction strength between two particles to be tuned
to essentially any value through application of an external magnetic field.
A detailed understanding of Feshbach resonances in 3D
underlies studies of, e.g., the BEC-BCS crossover
and $p$-wave pairing.
Confinement induced
resonances in 2D have been predicted~\cite{petr00,italian}, and 
recently been observed for $p$-wave 
interactions~\cite{gunter}, 
thus allowing the effective 2D coupling constant to be tuned 
to essentially any value including zero and infinity.
Assuming a strictly 2D geometry,
we propose a coupled channel model for the lowest partial
wave, i.e., $m=0$,
with four parameters, the scattering lengths 
$b_1$ and $b_2$ of the
``open'' and ``closed'' channel, the coupling strength $\beta$ and
the detuning $\epsilon$.
While $b_1$, $b_2$ and $\beta$ are fixed for a given system,
varying the detuning $\epsilon$ corresponds to changing the strength 
of an external magnetic field.
To the best of our knowledge, no coupled-channel
model for 2D resonances has been proposed to date.
Our model predicts that the dependence of the effective 2D scattering
length on the external control parameter is distinctly different than that
of the effective
3D scattering length, thus highlighting the non-intuitive
behavior
of systems with reduced dimensionality.
We also determine the binding energy and the occupation
of the open and closed channel across a 2D resonance.
These observables have been measured across a 3D 
resonance~\cite{observable_experiment1,observable_experiment2}
but not yet across a 2D resonance; however, we expect
measurements on quasi-2D systems to be performed in the near future.

To derive 
2D ZR potentials for 
any partial wave
we assume that the atom-atom potential $V_{int}(\rho)$ 
depends only on the distance
$\rho$ between the two atoms, that is, we neglect any
angular dependence that may arise from
spin-dependent interactions.
The center of mass motion can then be separated off, and 
the 
radial Schr\"odinger equation reads
\begin{eqnarray}
\label{eq_se}
\left[ \frac{\partial^2}{\partial \rho^2}+ 
\frac{1}{\rho}\frac{\partial}{\partial \rho}-
\frac{m^2}{\rho^2}- 
\frac{2 \mu}{\hbar^2} 
\left(V(\rho) - E \right)  \right] R_m(\rho)=0,
\end{eqnarray}
where $\mu$ denotes the reduced mass of the two-atom system,
and $V(\rho)=V_{int}(\rho)+V_{ext}(\rho)$ 
[$V_{ext}(\rho)$
denotes the $\rho$-dependent part
of an external confining potential, see below; for now we set $V_{ext}=0$].
In Eq.~(\ref{eq_se}), 
$m$ denotes the orbital quantum number; $m=0$ applies to
scattering between two 2D bosons or two 2D fermions with
opposite spin, $m=1$ to scattering between two spin-polarized
fermions, and so on. 
We now
derive $m$-dependent 
2D ZR pseudo-potentials, $V_{int}(\rho)=V^{ps}_m(\rho)$, which reproduce the
low-energy observables of a shape-dependent short-range
atom-atom
potential.

We 
write the pseudo-potential $V^{ps}_m(\rho)$ in terms of a 
$\delta$-shell of radius $s$ and a yet to be determined operator
$\hat{O}_m(\rho)$~\cite{stoc05},
\begin{eqnarray}
\label{eq_ps}
V^{ps}_m(\rho) = { \{ \hat{O}_m(\rho) \delta (\rho-s) \} }_{s\rightarrow 0}.
\end{eqnarray}
The solutions to Eq.~(\ref{eq_se})
for $V_{int} = V_m^{ps}$ 
can be written in terms of the 
cylindrical Bessel functions $J_m(k \rho)$ and $N_m(k \rho)$.
For $\rho < s$, only $J_m$, which is regular at the origin, contributes,
\begin{eqnarray}
\label{eq_inner}
R^-_m(\rho)=B_m J_m (k \rho),
\end{eqnarray}
where $k = \sqrt{2 \mu E / \hbar^2}$.
The $\delta$-shell introduces a 
phase
shift $\delta_m(k)$ of the $m$th partial wave so that 
the wave function for $\rho > s$
is  given by
\begin{eqnarray}
\label{eq_outer}
R^+_m(\rho)= A_m \left[ J_m (k \rho) -\tan (\delta_m(k)) N_m (k \rho) 
\right].
\end{eqnarray}
In Eqs.~(\ref{eq_inner}) and (\ref{eq_outer}), $B_m$ and $A_m$ denote
constants to be determined below.

For $s$-wave scattering ($m=0$),
the phase shift $\delta_0(k)$ determines the 
2D energy-dependent scattering length $a_0(k)$~\cite{verhaar},
\begin{eqnarray}
\label{eq_a2d0}
a_0(k) = \frac{2}{k}
\exp \left[ \frac{\pi}{2}
\cot \delta_0(k) - \gamma \right],
\end{eqnarray}
where $\gamma$ denotes Euler's constant.
The energy-dependent scattering length $a_0(k)$, which is always greater or
equal to zero, is defined such that the scattering wave function
has a node at $\rho=a_0(k)$. The unusual functional form of $a_0(k)$, 
i.e., the exponential
dependence on the phase shift,
is a direct consequence of the logarithmic dependence of $N_m(k \rho)$ 
on $k \rho$,
$N_0 (k \rho) \approx \frac{2}{\pi}[\ln(k \rho/2) + \gamma]$
for $k \rho \rightarrow 0$.
For higher partial waves, we define 
generalized energy-dependent ``scattering lengths''
$a_m(k)$, which have dimensions of (length)$^{2m}$, as
\begin{eqnarray}
\label{eq_a2dm}
a_m(k) = - 
\frac{\tan(\delta_m(k))}{k^{2m}}
\frac{\Gamma(m) \Gamma(m+1) 2^{2m}}{\pi} .
\end{eqnarray}
Since $a_1(k)$ has dimensions of (length)$^2$ we refer to it as 
scattering area.
Energy-independent generalized scattering lengths 
$a_m$ are readily defined through $a_m = \lim_{k \rightarrow 0} a_m(k)$.

Imposing continuity of the
wave function $R_m(\rho)$ at $\rho=s$,
that is, requiring $R_m^+(s)= R_m^-(s)$,
allows $B_m$ to be expressed in terms of $A_m$.
Integrating the Schr\"odinger equation from 
$ \rho=s- \epsilon$ to $s+\epsilon $, and then taking 
the limit $ \epsilon \rightarrow 0 $, results in
\begin{eqnarray}
\label{eq_deriv}
\frac{\hbar^2}{2 \mu}
\left[
\frac{\partial}{\partial \rho}R^+_m(\rho)-
\frac{\partial}{\partial \rho}R^-_m( \rho) \right]_{\rho = s} = 
\hat{O}_m(s)R_m( s).
\end{eqnarray}
Plugging Eqs.~(\ref{eq_inner}) and (\ref{eq_outer}) into 
Eq.~(\ref{eq_deriv}) and taking
$k s \ll 1$,
determines the operator $\hat{O}_m(s)$, and hence the 
pseudo-potential $V_m^{ps}(\rho)$.
For $m=0$, we find
\begin{eqnarray}
V_0^{ps}(\rho)=
\left\{  
\frac{-\frac{\hbar^2}{\mu} \tan (\delta_0(k))}
     {\left(1- \frac{2 \tan (\delta_0(k))}{\pi}f_0(k,\rho) \right) \pi \rho}
\frac{\partial}{\partial \rho}\rho 
\delta(\rho - s)  \right\}_{s \rightarrow 0},
\label{eq_pseudom0}
\end{eqnarray} 
where
$f_0(k, \rho) = 1 + \gamma + \ln(k \rho/2)$.
The explicit $k$ dependence  of the $m=0$ ZR potential drops out when 
$V_0^{ps}$ is written in terms $a_0(k)$.
Our $s$-wave pseudo-potential agrees with the $\Lambda$
potential derived in 
Ref.~\cite{olsh02} if one sets $\Lambda$ equal to 
$k$ (see also Ref.~\cite{wolsch}).
Below, we use the boundary condition implied by the pseudo-potential
$V_0^{ps}$ to develop
a 
coupled-channel ZR model
for 2D scattering.

For $m>0$, a straightforward yet somewhat tedious calculation gives
\begin{eqnarray}
\label{eq_pseudom}
V_m^{ps}(\rho,k)=\nonumber \\
\left\{ 
\frac{- \frac{\hbar^2 [\Gamma(m + 1)]^2 \tan (\delta_m(k))}{\mu ( 2m)! \pi }
\left(\frac{2}{k} \right)^{2m} }
{(1 + \frac{\tan (\delta_m(k))}{\pi}
f_m(k, \rho))\rho^{m+1}
}  
\frac{\partial^{2m}}{\partial \rho^{2m}}
\rho^m \delta(\rho -s) \right\}_{s \rightarrow 0},
\end{eqnarray}
where
\begin{eqnarray}
f_m(k, \rho) = 
\bar{\psi}(m) 
 -2 \ln \left(\frac{k \rho}{2} \right)-
\sum_{r=0}^{r=2m-1} 
\frac{2}{2m-r}.
\end{eqnarray}
Here, $\bar{\psi}(m)=\psi(1) + \psi(m+1)$, 
where $\psi$ denotes the digamma function.
As written in Eq.~(\ref{eq_pseudom}), the $m>0$
pseudo-potential leads, despite the 
regularization operator, to divergencies at $\rho \rightarrow 0$
if the 
energy-dependent coefficients $C_m$ and $D_m$ of the $\rho^{-m}$ 
and $\ln(\rho)\rho^m$ terms in the expansion of 
the eigenfunction sought
differ from the corresponding 
coefficients $c_m$ and $d_m$ of the expansion of the 
irregular 
free particle solution
$N_m(k \rho)$. To cure this divergence, the right hand side of
Eq.~(\ref{eq_pseudom})
has to be multiplied
by $d_mC_m/(c_mD_m)$,
resulting in
a pseudo-potential that 
has to be determined self-consistently~\cite{tobepublished}.
A similar self-consistency condition
is not needed for systems with odd dimensionality.

The bound state energies $E_m^{bind}$ of the
pseudo-potentials $V_m^{ps}$ can be determined through analytic 
continuation~\cite{stoc05}.
For $m=0$, we recover the well known expression for the
ZR binding energy $E_0^{bind}$~\cite{jens04},
\begin{eqnarray}
\label{eq_bind0}
E_0^{bind}= \frac{-\hbar^2}{2 \mu (a_0(k))^2} \; 4 \exp(-2 \gamma).
\end{eqnarray}
For $m>0$, the
binding energies $E_m^{bind}$ are given
by
\begin{eqnarray}
\label{eq_bindm}
E_m^{bind}= \frac{ -2 \hbar^2}{\mu [-i (-1)^m a_m(k)]^{1/m}}
\;  \left( \frac{ \Gamma (m)\Gamma(m+1)}{\pi} \right)^{1/m}.
\end{eqnarray}
The occurance of an ``$i$'' in Eq.~(\ref{eq_bindm}) appears odd
at first sight.
This puzzle is resolved by 
noting that the 2D generalized energy-dependent scattering lengths $a_m(k)$
with $m>0$
are, in contrast to the 1D and 3D counterparts, complex.
Consequently, the binding 
energies $E_m^{bind}$ are determined by those $a_m(k)$
for which the real part vanishes.
For $m=1$, e.g., 
we find excellent agreement between the exact binding energy 
determined from the 
energy-dependent scattering area via Eq.~(\ref{eq_bindm}) and that for
a square well potential with range $\rho_0$,
$V_{int}(\rho)=V_{SW}(\rho)= -V_0$ 
for $\rho < \rho_0$ and zero otherwise.

We now use the proposed 
pseudo-potentials 
to determine
the eigenspectrum of two atoms in 2D under
external harmonic confinement,
that is, in Eq.~(\ref{eq_se}) we consider 
$V_{ext}(\rho) = \mu \omega^2 \rho^2/2$.
Such a two-body system can be realized experimentally
with the aid of a 1D optical lattice with doubly-occupied lattice sites,
for which the tunneling between neighboring sites is neglegible.
The solutions for $\rho < s$ and $\rho > s$ are proportional to the
confluent hypergeometric
functions $M$ and $U$, i.e.,
$R_m^-(\rho) \propto \rho^m \exp(-\rho^2/2a_{ho}^2)
M(-\chi, m+1, \rho^2/a_{ho}^2)$
and
$R_m^+(\rho) \propto \rho^m \exp(-\rho^2/2a_{ho}^2)
U(-\chi, m+1, \rho^2/a_{ho}^2)$,
where $a_{ho}=\sqrt{\hbar/(\mu \omega)}$
denotes the oscillator length and $\chi$ a non-integer quantum number,
$E_{m \chi}=(2 \chi + 1 + m) \hbar \omega$.
Enforcing continuity at $\rho = s$ determines the eigenenergies
$E_{m \chi}$ 
implicitly in terms of $a_m(k)$,
\begin{eqnarray}
\label{eq_psboxm}
(-1)^{m+1} \frac{\Gamma(-\chi-m)}{\Gamma(-\chi)} 
 \sum_{r=0}^{m-1} \frac{(-\chi-m)_r (-1)^{m-r}}{ (1-m)_r r! (2m-2r)!!} \nonumber \\
 = \frac{\ln \left(\chi+\frac{m+1}{2} \right) -\psi(-\chi)}
{\Gamma(m) \Gamma(m+1)}
+
\frac{a_{ho}^{2m}}{a_m(k_{m \chi})} \frac{1}{\left(\chi+\frac{m+1}{2}\right)^m} ,
\end{eqnarray}
where
$k_{m \chi}=\sqrt{2 \mu E_{m \chi}}/\hbar$ and
$(x)_r = x(x+1)\cdots (x+r-1)$ with $(x)_0=1$. 
Equation~(\ref{eq_psboxm}) contains the 
generalized {\em{energy-dependent}}~\cite{blum02} 
2D scattering length
$a_m(k)$ evaluated at $k=k_{m \chi}$ and is valid for 
$m>0$;
for $m=0$ 
the eigenenergies are given by
Eq.~(\ref{eq_ccenm0}) with $\beta=0$ and $b_1=a_0(k)$.

\begin{figure}[tbp]
\vspace*{-1.8in}
\centerline{\epsfxsize=3.2in\epsfbox{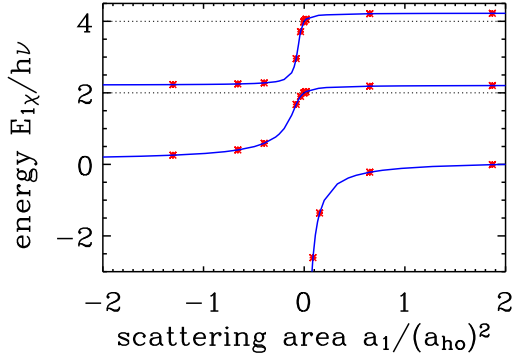}}
\vspace*{-.2in}
\caption{
$E_{1 \chi}$ 
as a function of $a_1(k=0)$ ($\omega=2 \pi \nu$).
Asterisks show 
$E_{1 \chi}$, Eq.~(\ref{eq_psboxm}), for the energy-dependent 
pseudo-potential $V_1^{ps}(\rho)$ and solid lines 
show $E_{1 \chi}$
for $V_{SW}$ with $\rho_0=0.01 a_{ho}$.
For comparison, 
horizontal dotted lines indicate the energy levels for $a_1=0$.
}
\label{fig2}
\end{figure}

Asterisks in Fig.~\ref{fig2}
show the $m=1$ eigenenergies for two particles interacting 
through the proposed ZR potential, Eq.~(\ref{eq_psboxm}), 
under harmonic confinement
over a large range of 
zero-energy scattering areas $a_1$. 
For comparison, solid lines show the exact eigenenergies,
determined semi-analytically, for two
particles under harmonic confinement 
interacting through a square well potential with range 
$\rho_0=0.01 a_{ho}$.  
Figure~\ref{fig2} illustrates 
excellent
agreement between the
eigenenergies for two 2D particles 
interacting 
through the pseudo-potential and those interacting through the
square well potential.
We find
similar behaviors for higher partial waves.

To describe the two-body physics across a 2D Feshbach resonance
for two bosons or for two fermions with opposite spin,
we develop a coupled-channel ZR model for $m=0$. 
For $\rho>0$, the wave function with components
$R_0^{(1)}(\rho)$ and $R_0^{(2)}(\rho)$ satisfies the
harmonic oscillator Schr\"odinger equation in the relative coordinate,
\begin{eqnarray}
\label{eq_cc}
\left(
	\begin{array}{c} 
	\left[ \frac{\hbar^2}{2 \mu} ( \frac{\partial^2}{\partial \rho^2}+ 
\frac{1}{\rho}\frac{\partial}{\partial \rho})
-\frac{1}{2} \mu \omega^2 \rho^2
+ E \right] R_0^{(1)}(\rho)  \\
	\left[ \frac{\hbar^2}{2 \mu} ( \frac{\partial^2}{\partial \rho^2}+ 
\frac{1}{\rho}\frac{\partial}{\partial \rho})
-\frac{1}{2} \mu \omega^2 \rho^2
+ E - \epsilon \right] R_0^{(2)}(\rho)
	\end{array} 
\right)
=
0,
\end{eqnarray}
where $\epsilon$ denotes a detuning ($\epsilon \ge 0)$, which can be changed,
e.g.,
by varying 
the strength of an  external magnetic field.
The 2D  
pseudo-potential 
$V_{0}^{ps}$ imposes a boundary condition at $\rho=0$, 
which we parameterize as
\begin{eqnarray}
\label{eq_bcm0}
\left(
	\begin{array}{cc} 
	\ln(\rho/b_1) & \beta \\
	\beta & \ln(\rho/b_2) 
	\end{array} 
\right)
\left(
	\begin{array}{c} 
	 \rho \frac{\partial}{\partial \rho} R_0^{(1)} \\
	 \rho \frac{\partial}{\partial \rho} R_0^{(2)}
	\end{array}
\right) 
= 
\left(
	\begin{array}{c} 
	 R_0^{(1)} \\
	 R_0^{(2)} 
	\end{array}
\right).
\end{eqnarray}
The coupling between the two channels is characterized by the
dimensionless parameter $\beta$.
To ensure a divergence-free treatment,
Eq.~(\ref{eq_bcm0}) takes this coupling parameter
to be proportional to the {\em{derivative}} of the wave function 
components, and
not, as done in 3D~\cite{dunn05}, to be proportional
to the wave function components themselves.
The scattering length 
$a_{0}^{CC}(E,\epsilon)$
predicted by Eqs.~(\ref{eq_cc}) 
and (\ref{eq_bcm0}) in the limit $\omega \rightarrow 0$  is
\begin{eqnarray}
\label{eq_ccm0sc}
a_{0}^{CC}(E,\epsilon) = b_1 \exp \left\{ -\beta^2/ \left[
\gamma+ \frac{1}{2} \ln \left( \frac{\mu b_2^2}{2 \hbar^2} (\epsilon-E)
\right) 
\right] \right\}.
\end{eqnarray}

\begin{figure}[tbp]
\vspace*{-1.8in}
\centerline{\epsfxsize=3.2in\epsfbox{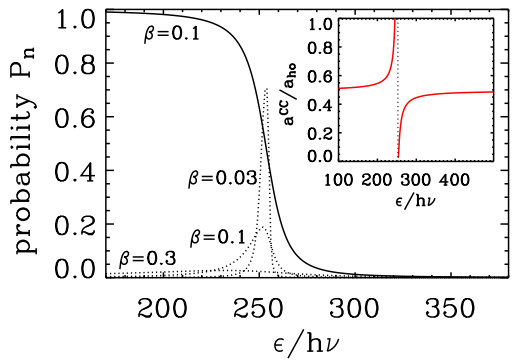}}
\vspace*{-.2in}
\caption{
Molecular fractions $P_0$ (solid line) and  $P_1$ (dotted lines),
Eq.~(\protect\ref{eq_molfrac}),
for $b_1=0.5 a_{ho}$, $b_2=0.05 a_{ho}$ and different
coupling constants $\beta$ 
as a function of $\epsilon$.
Inset:
2D scattering length $a_0^{CC}(E, \epsilon)$ 
as a function of $\epsilon$ 
for
$E = 0$, $b_1=0.5 a_{ho}$, $b_2=0.05 a_{ho}$ and $\beta=0.1$.}
\label{fig3}
\end{figure}

To determine the behavior of the 
scattering length $a_0^{CC}(E,\epsilon)$ in the vicinity
of the resonance as a function of the magnetic field strength $B$, we
Taylor-expand $a_0^{CC}(E,\epsilon)$ about the resonance position
$\epsilon_R$, which is given by
the binding energy $E_0^{bind}$ of the strongly closed 
molecular channel, $\epsilon_R=2 \exp(-2 \gamma)\hbar^2/(\mu b_2^2)$.
We find a simple functional form for $a_0^{CC}(E=0,B)$
in terms of the background scattering length
$A_{bg}$, the resonance width $\Delta$ and the resonance
position $B_R$,
\begin{eqnarray}
\label{eq_ccm0sc_b}
a_0^{CC}(E=0,B)= A_{bg} \exp [ - \Delta/(B-B_R) ],
\end{eqnarray}
where
$A_{bg}=b_1$, $B_R=\epsilon_R/M$ and $\Delta=2 \beta^2 \epsilon_R/M$
($M$ denotes the difference in magnetic
moment between atoms
in the open and closed channel).
Just as the 2D single-channel scattering length $a_0$ [Eq.~(\ref{eq_a2d0})], 
the functional dependence of the 2D coupled-channel
scattering length $a_0^{CC}$ on $B$
near a resonance is distinctly different from that of the 3D
counterpart.
In principle,
the parameters $b_1$, $b_2$ and $\beta$ can be determined by comparing 
Eq.~(\ref{eq_ccm0sc_b}) with experimental data for 
a specific 2D Feshbach resonance. Since
no such data exist to date,
the inset of Fig.~\ref{fig3}
illustrates the behavior of $a_0^{CC}$ as a function of $\epsilon$ 
for $E=0$,
$a_1=0.5 a_{ho}$, $a_2=0.05 a_{ho}$ and $\beta=0.1$.
The scattering length $a_0^{CC}$ changes from infinity
to zero at the resonance value $\epsilon_R$ (indicated by a vertical 
dotted line).

\begin{figure}[tbp]
\vspace*{-1.8in}
\centerline{\epsfxsize=3.2in\epsfbox{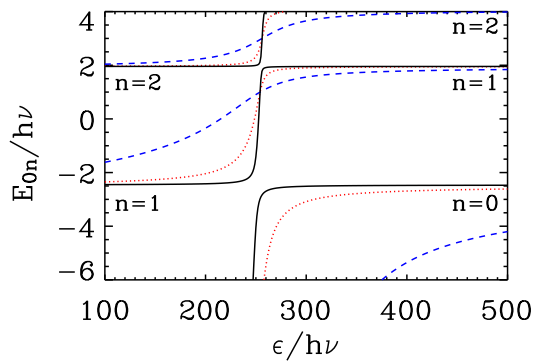}}
\vspace*{-.2in}
\caption{
Eigenenergies $E_{0n}$, Eq.~(\ref{eq_ccenm0}),
for
$b_1=0.5 a_{ho}$, $b_2=0.05 a_{ho}$ and three different values of
$\beta$, i.e., $\beta=0.03$ (solid lines), $0.1$ (dotted lines)
and $0.3$ (dashed lines)
as a function of $\epsilon$.}
\label{fig4}
\end{figure}

We find an implicit eigenequation
for the eigenenergies $E_{0n}$, $E_{0n} < \epsilon$, 
of the coupled-channel ZR model
for two particles under harmonic confinement with $m=0$
[Eqs.~(\ref{eq_cc}) and (\ref{eq_bcm0})], 
\begin{eqnarray}
\label{eq_ccenm0}
\beta^2 \left\{
\ln \left(\frac{b_1}{a_{ho}} \right) 
+ \frac{1}{2} 
\psi \left(\frac{1}{2} - \frac{E_{0n}}{2 \hbar \omega} \right) - \psi(1) 
\right\}^{-1} \nonumber \\
= \left\{
\ln \left(\frac{b_2}{a_{ho}} \right) 
+ \frac{1}{2} \psi \left(\frac{1}{2} - \frac{E_{0n}-\epsilon}{2 \hbar \omega} \right) -  \psi(1)  
\right\}.
\end{eqnarray}
Figure~\ref{fig4} 
shows the eigenenergies $E_{0n}$ as a function of $\epsilon$
for $b_1=0.5 a_{ho}$, $b_2=0.05 a_{ho}$ and three different values
of $\beta$, i.e., $\beta=0.03$ (solid lines),
$0.1$ (dotted lines) and $\beta=0.3$ (dashed lines).
The non-vanishing coupling leads to a series of energy level crossings 
at $\epsilon \approx \epsilon_R$, 
which become narrower as the coupling $\beta$
decreases.
For higher-lying states, i.e., larger $E_{0n}$, 
the resonance of the energy-dependent 
scattering length $a_0^{CC}(\epsilon,E)$, Eq.~(\ref{eq_ccm0sc}),
moves to larger $\epsilon$ values, and thus explains why the 
energy level crossings move to larger $\epsilon$ for higher $n$.
The energy of the state $n=1$ for $\epsilon < \epsilon_R$
corresponds to the binding energy of
the open channel, i.e., of channel $(1)$.
We checked that a 2D coupled channel square-well system
(see Ref.~\cite{greene} for the 3D analog)
reproduces the results obtained for our proposed coupled-channel ZR system.

Our coupled channel model leads to a mixing of the 
strongly closed molecular channel $R_{0n}^{(2)}$ and the
open channel $R_{0n}^{(1)}$. 
To quantify the admixture of the strongly closed molecular level
across the resonance, we define the molecular fraction $P_n$~\cite{greene},
\begin{eqnarray}
\label{eq_molfrac}
P_n =
\int |R_{0n}^{(2)}|^2 \rho d\rho / \left[
\int \left( |R_{0n}^{(1)}|^2 + |R_{0n}^{(2)}|^2 \right)
\rho d\rho \right].
\end{eqnarray}
The main part of Fig.~\ref{fig3} shows $P_n$ 
for $b_1=0.5 a_{ho}$, $b_2=0.05 a_{ho}$ and different $\beta$
values 
as a function of $\epsilon$;
a solid line shows $P_0$ 
for $\beta=0.1$,
and dotted lines show $P_1$ for
$\beta=0.03$, $0.1$ and $0.3$.
The molecular fraction $P_0$ is close to one for small
$\epsilon$, and drops to zero as $\epsilon$ is swept across the resonance.
This indicates that the character of the $n=0$ state changes from 
``strongly closed
molecular'' to ``weakly closed molecular'' as $\epsilon$ changes from
$\epsilon < \epsilon_R$ to
$\epsilon > \epsilon_R$.
The molecular fraction of the $n=1$ state is close to zero away
from resonance for all coupling strengths
considered. Near resonance, however, $P_1$ depends on $\beta$.
For weak coupling,
$P_1$ approaches
one on resonance. For strong coupling, in contrast, $P_1$
is comparatively small as shown in Fig.~\ref{fig3} for $\beta=0.3$.
This suggests that 
the 2D analog of the BEC-BCS crossover can be best studied 
utilizing 2D Feshbach resonances with strong coupling  
for which the admixture of the
strongly closed molecular channel is small.
Similar behavior is found for 3D systems~\cite{moor05}.

In summary,
this paper derives a series of 2D pseudo-potentials,
which 
describe the low-energy scattering of two particles
with partial wave $m$.
The boundary condition implied by the $m=0$ pseudo-potential
is then used to
develop an analytically
solvable coupled-channel
model, which describes
the physics across a 
2D Feshbach resonance.
The predicted dependence of the effective 2D scattering length
$a_0^{CC}$ on the strength $B$ of the external magnetic field,
Eq.~(\ref{eq_ccm0sc_b}),
may prove useful in analyzing experimental data for quasi-2D 
Bose gases or two-component Fermi gases.
We also determine the binding energy of 2D dimers,
which can be measured with present-day technology by utilizing optical lattices
with doubly-occupied lattice sites in the no-tunneling
regime~\cite{observable_experiment1},  across a Feshbach resonance.

Fruitful discussions with 
Chris Greene 
and Stefano Giorgini,
 and
support by the NSF under grant PHY-0331529
are gratefully acknowledged.


\end{document}